 \pdfoutput=1

\documentclass[10pt]{IEEEtran}
\PassOptionsToPackage{bookmarks={false}}{hyperref}

\usepackage{nopageno}

\usepackage{multirow}
\usepackage{amsmath}
\usepackage{float}

\usepackage{booktabs}
\usepackage{tabularx}
\usepackage{ragged2e}
\usepackage{color}
\usepackage[utf8]{inputenc}
\usepackage{tabularx}
\usepackage{capt-of}
\usepackage[algo2e]{algorithm2e}
\usepackage{lipsum,multicol}
\usepackage{array,multirow}

\PassOptionsToPackage{bookmarks={false}}{hyperref}
\usepackage{algorithm}
\usepackage{caption}
\usepackage{subcaption}
\usepackage{amsmath}
\usepackage{amsthm}
\usepackage{amssymb}
\usepackage{fancyvrb}
\usepackage{algorithmic}
\usepackage[T1]{fontenc}
\usepackage[scaled]{helvet}
\usepackage{lscape}
\usepackage[dvips,final]{graphicx}
\graphicspath{{./}{./figures/}}
\usepackage{enumerate}
\usepackage{epstopdf}
\usepackage{float}
\usepackage{url}
\usepackage{caption}
\usepackage{booktabs}
\usepackage{cite}
\usepackage{array}
\usepackage{tabularx}
\def\BibTeX{{\rm B\kern-.05em{\sc i\kern-.025em b}\kern-.08em T\kern-.1667em\lower.7ex\hbox{E}\kern-.125emX}}

\newcommand{\comment}[1]{ }

\IEEEoverridecommandlockouts
\begin{document}
\newcommand{\proposed}{{\tt Bert}}

\title{Bert: Scalable Source Routed Multicast for Cloud Data Centers}


\author{Jarallah Alqahtani, Bechir Hamdaoui ~\\
\small Oregon State University, \small Corvallis,\\
\small \{alqahtaj,hamdaoui\}@eecs.oregonstate.edu~

}


\maketitle 
\begin{abstract}
Traditional IP multicast routing is not suitable for cloud data center (DC) networks due to the need for supporting large numbers of groups with large group sizes. State-of-the-art DC multicast routing approaches aim to overcome the scalability issues by, for instance, taking advantage of the symmetry of DC topologies and the programmability of DC switches to compactly encode multicast group information inside packets, thereby reducing the overhead resulting from the need to store the states of flows at the network switches. However, although these scale well with the number of multicast groups, they do not do so with group sizes, and as a result, they yield substantial traffic control overhead and network congestion.
In this paper, we present \proposed, a scalable, source-initiated DC multicast routing approach that scales well with both the number and the size of multicast groups, and does so through clustering, by dividing the members of the multicast group into a set of clusters with each cluster employing its own forwarding rules. Compared to the state-of-the-art approach, \proposed~yields much lesser traffic control overhead by significantly reducing the packet header sizes and the number of extra packet transmissions, resulting from the need for compacting forwarding rules across the switches.

\end{abstract}
\thispagestyle{empty}


\section{\sc {Introduction}} \label{sec:Introducation}
Today's cloud data centers (DCs) host hundreds of thousands of tenants~\cite{tenants}, with each tenant possibly running hundreds of workloads supported through thousands of virtual machines (VMs) running on different servers~\cite{Azure,Google,AWS}. These workloads often involve one-to-many communications among the different servers as required by the supported applications~\cite{shvachko2010hadoop,dean2008mapreduce}.
%
%
Therefore, to enable efficient communication and data transfer among the different servers running VMs supporting the same workload/application, multicast routing protocol designs need to be revisited to suit today's cloud data center network topologies.
Traditional IP multicast routing is primarily designed for arbitrary network topologies and Internet traffic, with focus on reducing CPU and network bandwidth overheads, and hence is not suitable for DCs due to the need for supporting large numbers of groups in commodity switches with limited memory capability. In other words, DC switches will have to maintain per-group routing rules for all multicast addresses, because they cannot be aggregated on per prefix basis.

That is said, there have been few research efforts devoted to overcome this scalability issue~\cite{avalanche,ipms,miniforest,Elmo,li2011rdcm,jia2013scalable,cui2014dual}. For instance, Elmo~\cite{Elmo}, a recently proposed source-initiated multicast routing approach for DCs, overcomes the scalability issue and is shown to support millions of multicast groups with reasonable overhead in terms of switch state and network traffic. Elmo does so by taking advantage of programmable switches~\cite{PS} and the symmetry of DC topologies to compactly encode multicast group information inside packets, thereby reducing the overhead resulting from the need to store the states of flows at the network switches.
However, although Elmo scales well with the number of multicast groups, it does not do so with multicast group sizes. When considering large multicast group sizes, Elmo header can carry on several hundreds of bytes extra, which increases traffic overhead in the network. In addition, the number of extra transmissions Elmo incurs due to compacting of packet rules increases significantly with the size of multicast group, yielding higher traffic congestion in the DC's downlinks. To overcome Elmo's aforementioned limitations, we propose in this paper \proposed, a source-initiated multicast routing for DCs. Unlike Elmo, \proposed~scales well with both the number and the size of multicast groups, and does so through clustering, by dividing the members of the multicast group into a set of clusters with each cluster employing its own forwarding rules. In essence,
\proposed~yields much lesser multicast traffic overhead than Elmo by significantly reducing (1) the forwarding header sizes of multicast packets and (2) the number of per group member transmissions resulting from the need for compacting forwarding rules across the switches.


The rest of this paper is organized as follows. We briefly illustrate the network architecture of modern DCs, and describe the limitations of prior related  state-of-the art works in Section~\ref{sec:LimititionOfTheStateOfTheArt}.  We present the proposed multicast routing scheme, \proposed, in Section \ref{sec:TheProposedTechnique}. We study and evaluate the performances of \proposed~and compare them to those obtained under Elmo in Section~\ref{sec:Evaluation}. We conclude the paper in Section~\ref{sec:Conclusion}.


%
%
%


\section{\sc {Limitations of the State of the Art}}
\label{sec:LimititionOfTheStateOfTheArt}
\begin{figure*}[ht]
   \centering
\begin{subfigure}[b]{0.55\textwidth}
   \centering
   \hspace*{-2cm}
   \includegraphics[height=7cm, width=8cm]{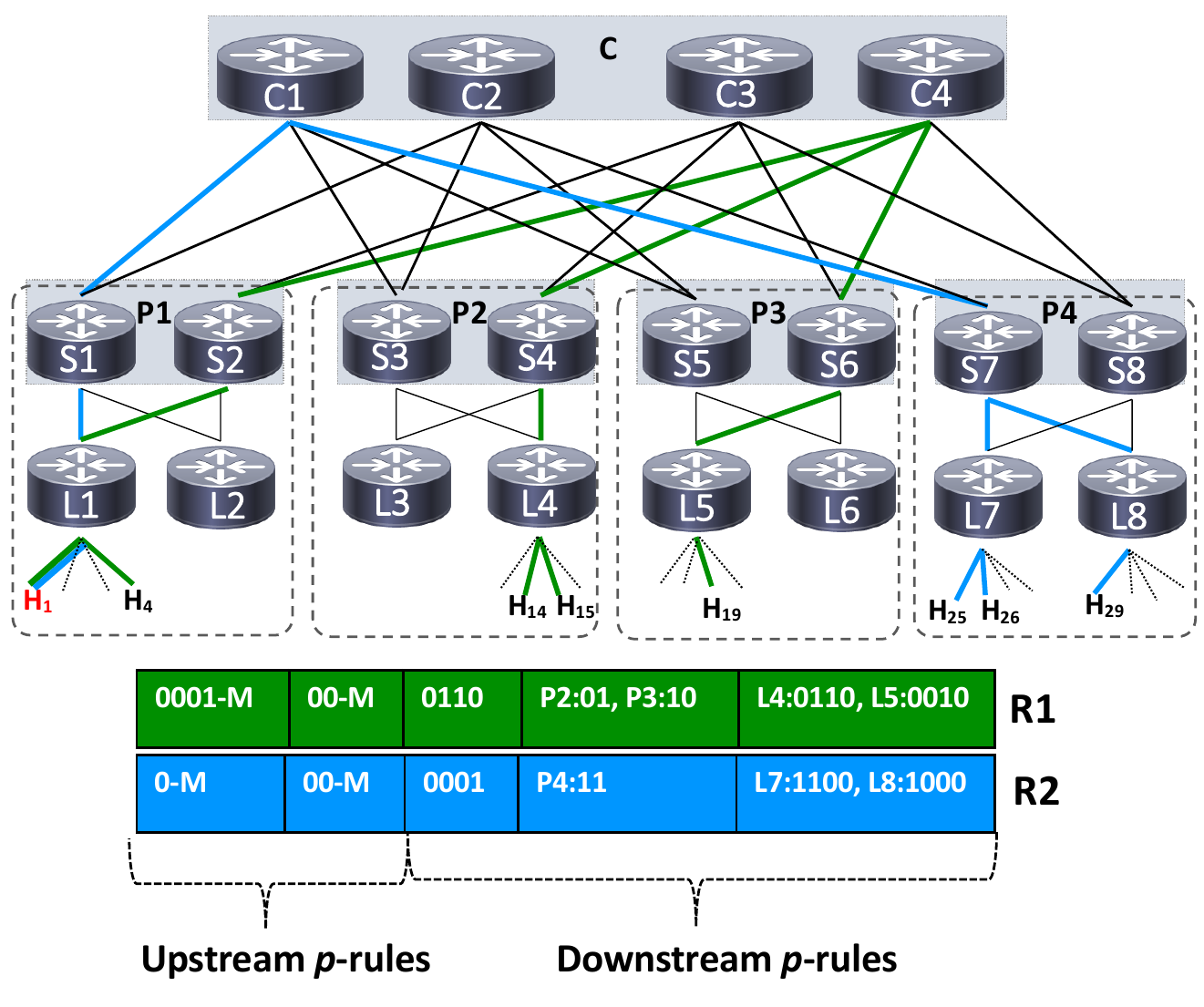}
   \caption{\proposed}
   \label{fig:fig1a}
\end{subfigure}%
\begin{subfigure}[b]{0.55\textwidth}
   \centering
   \hspace*{-1cm}
   \includegraphics[height=7cm, width=8cm]{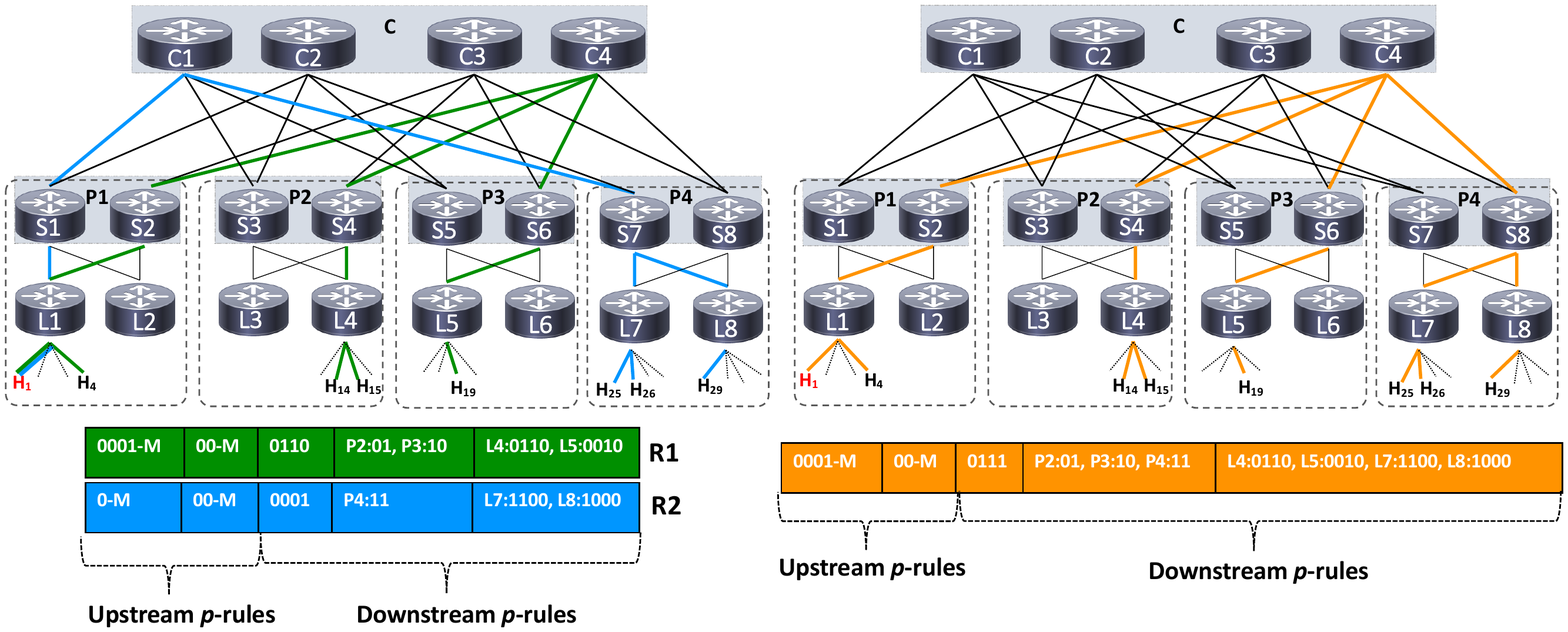}
   \caption{Elmo}
   \label{fig:fig1b}
\end{subfigure}%
   \caption{ An example of multicast tree on a three-tier Clos topology with four pods. In this topology, there are 4 hosts under each leaf switch (Top of Rack). $H_1$ is the source of multicast groups, and $H_4$, $H_{14}$, $H_{15}$, $H_{19}$, $H_{25}$, $H_{26}$, and $H_{29}$ are the receivers of multicas group.}
   \label{fig:fig1}
\end{figure*}


\subsubsection{Background---DC Topologies}
Large-scale DCs typically are multi-rooted tree-based topologies (e.g., fat-tree~\cite{fat-tree} and its variants~\cite{vl2,f10,cft}). These types of topologies provide large numbers of parallel paths to support high bandwidth, low latency, and non-blocking connectivity among servers.
The servers are tree leaves, which are connected to top-of-rack (ToR) (edge/leaf) switches.
In general, DCs contain three types of switches, leaf, spine, and core, with each type residing in one layer, as shown in Fig.~\ref{fig:fig1}. At the lowest layer, leaf (aka edge) switches are interconnected through spine (aka aggregation) switches, which constitute the second layer of switches. The core switches, constituting the top/root layer, serve as connections among the spine switches. With such a DC topology, every server can communicate with any other server using the same number of hops.

\subsubsection{Multicast in DCs}



DC multicast has been studied from a different point of views. For example, frameworks proposed in~\cite{mc_vne1, mc_vne2} studied the resource allocation and embedding of multicast virtual networks. Mainly they focused on how to place and restore VMs to provide high performance non-blocking multicast virtual networks while reducing hardware cost in Fat-Tree DCs.
Other works, including ours, focused on the scalability problem of multicast routing in DCs. These works relied either on decentralized protocols such as IGMP and PIM~\cite{miniforest} or centralized ones such as SDN-based approaches~\cite{avalanche, reliableli2013, ipms, cui2014dual}. Even though these approaches overcome scalability issue of multicast routing by supporting large numbers of multicast groups, they perform poorly when the size of groups are large.

\subsubsection{Elmo}
Elmo~\cite{Elmo} is a recently proposed DC multicast routing that scales well with number of multicast groups. Elmo is a source-based routing, which encodes packet forwarding state/rules in packet headers to limit flow state information that DC switches will have to maintain. It also exploits the programmable capability of the DC switches and the symmetry of DC topologies to compactly encode multicast group information inside packets, and thereby reduces packet header overhead, and consequently, network traffic load.

Even though Elmo is shown to scale well with the number of multicast groups, it still suffers from scalability issues in terms of incurred traffic overhead when it comes to large group sizes. For example, a packet header could be as large as $325$ bytes to contain all $p$-rules (packet rules)~\cite{Elmo}, incurring excessive network traffic overhead and link congestions. Elmo tries to overcome this by: (1) removing per-hop $p$-rules from the header as packets traverse the network switches; unfortunately, the downstream spine and leaf switches, which happen to consume most of the header space, are removed last, making most of the traffic overhead go over most of the network topology. (2) Switches in downstream paths with same or similar bitmaps are mapped to a single bitmap. For example, as shown in Fig.~\ref{fig:fig1a}, at the leaf layer, $L_7$ and $L_8$ can share one $p$-rule; e.g. $L_7,L_8: 1100$, yielding one extra transmission in $L_8$. However, sharing bitmaps results in extra packet transmissions, which they too increase traffic overhead.

In order to overcome the aforementioned challenges of Elmo, we propose \proposed, which first clusters the set of multicast destination members into multiple subsets/clusters, and then encodes multicast information in packet headers for each of these clusters. Our proposed multicast routing approach, \proposed, outperforms Elmo in terms of traffic overhead by significantly reducing (1) packet header sizes and (2) the number of extra transmissions resulting from the need for compacting forwarding rules.

\section{\sc {The Proposed Multicast Routing: \proposed}}
\label{sec:TheProposedTechnique}

\subsection{Motivating Example}
In this section, a detailed example is presented and illustrated in Fig.~\ref{fig:fig1} to explain the limitations of Elmo and motivate the design of the proposed scheme, \proposed.
At the high level, for each multicast group, the controller first computes a multicast tree and the forwarding rules, and then, installs these rules in the hypervisor of the multicast group source. The hypervisor intercepts each multicast packet and adds the forwarding rules to the packet header.
Elmo essentially focuses on how to efficiently encode a multicast forwarding policy in the packet header. Whereas \proposed, in addition to efficiently encoding the forwarding rules, aims to alleviate traffic overhead caused by header size and extra packet transmissions in the downstream paths.
The forwarding header consists of a succession of $p$-rules that include rules for upstream leaf and spine switches, as well as for the downstream core, spine, and leaf switches. Each switch in the multicast tree will remove its $p$-rules from the header when forwarding the packet to the next layer. For both Elmo and \proposed, each multicast packet's journey can be explained in two main phases:

\subsubsection{Upstream (leaf switches to core switches) path}
The $p$-rules for upstream switches (leaf and spine) consist of downstream ports and a multipath flag. When the packet arrives at the upstream leaf switch, the switch forwards it to the given downstream ports as well as multipathing it to the upstream spine switch using an underlying multipath routing scheme; e.g. ECMP~\cite{ECMP}.
In Elmo, only one packet goes through upstream paths.
Using Fig.~\ref{fig:fig1b} for illustration, leaf switch $L_1$ first removes its $p$-rules ($0001-M$) from the packet, then forwards it to the host $H_4$ as well as multipathing it to any spine switch $P_1$. The upstream spine switches will do the same to forward the packet to the core switches.
Our proposed \proposed, on the other hand, first clusters the destination members of the multicast group into multiple (two in the example) clusters, and then sends multiple (two in the example) copies of the packet (with different headers but same payload), one for each cluster; more detail on the clustering part will be provided later.
The first packet has the same upstream $p$-rules as Elmo; e.g. $R1$, while the second packet (e.g. $R2$) does not have any downstream rules for the leaf and spine switches to avoid any extra transmissions.
Even though duplicate packets will incur some minor traffic in upstream paths, it will reduce the traffic in the downstream paths substantially when compared to Elmo. That is, the overall traffic reduction in both the upstream and downstream will be significantly reduced under \proposed.

\subsubsection{Downstream (core switches to leaf switches) path}
The $p$-rules for the core, spine, and leaf switches in the downstream path consist of downstream ports and switch IDs. In the downstream path, the core switches forward the packet to the given pod based on the core switch $p$-rules. In Elmo, one core switch sends the packet to the spine switches, which in turn forward it (based on the spine switch $p$-rule) to the leaf switches. The leaf switches do the same to deliver the packet to the destination hosts. Note that because the topology symmetry, any core switch can forward the packet to the destination pods.
Referring to the example in Fig.~\ref{fig:fig1b} again, in Elmo, core switch $C$ sends the packet to $P2$, $P3$ and $P4$ switches (three packets in total), once the packet arrives at the downstream spine switch, it is then forwarded based on the spine switch $p$-rules to the leaf switches. These leaf switches do, in turn, the same to deliver the packet to the destination hosts. For example, when all leaf switches in the multicast tree share one $p$-rule---which should then be bitwise $OR$ of all these leaf switches (i.e., $L_4,L_5, L_7, L_8: 1111$), Elmo incurs 10 extra packet transmissions (see Fig.~\ref{fig:fig1b}).



Unlike Elmo, to reduce the number of extra unneeded transmissions, \proposed~first clusters the destination members into multiple (two in the example) clusters, and then sends a different copy for each cluster in the downstream (all copies have the same payload and size but different header/rules).
Referring to Fig~\ref{fig:fig1a} again for illustration, in \proposed, $C4$ forwards the first packet (e.g. $R1$) to $P2$ and $P3$, while $C1$ forwards the second packet (e.g. $R2$ ) to $P4$. Note that the number of core-pod packets, which is three in the example, is the same in both Elmo and \proposed. However, \proposed~reduces substantially the number of extra packet transmissions from leaf switch to end hosts. To illustrate, when, as done above for the case of Elmo, all leaf switches within the same cluster share one $p$-rule (i.e., in the example of Fig~\ref{fig:fig1a}, when each of $R1$ and $R2$ compacts its leaf switch rules in one rule only, with $R1$'s and $R2$'s rules becoming $L_4,L_5: 0110$ and $L_7,L_8: 1100$ respectively), \proposed~incurs only 2 extra packet transmissions as opposed to 10 in the case of Elmo.
Taking into account both the upstream and downstream paths, compared to Elmo, \proposed~incurs 3 more extra transmissions in the upstream (1 extra in each upstream layer), but 8 lesser transmissions in the downstream, thereby reducing the total extra transmissions by 5 compared to Elmo.


In addition to reducing the extra packet transmissions, the header size for the downstream packet in \proposed~is reduced.  For example, the header size of the first packet ($R1$) is 36 bits and that of the second packet ($R2$) is 21 bits. To identify switches, we use three bits for each of the spine and leaf switches. Hence, the average header size in \proposed~is about 29 bits per packet. Elmo packet header, on the other hand, is of sizes 55 bits (see Fig.~\ref{fig:fig1b}).
In general, the average header size for the downstream packet in \proposed~is $\frac{1}{k}$ of that of Elmo's packet, where $k$ is the number of the clusters of the multicast group, a design parameter of \proposed.

\begin{figure*}
   \centering
\begin{subfigure}[b]{0.55\textwidth}
   \centering
   \hspace*{-2cm}
   \includegraphics[height=7cm, width=8cm]{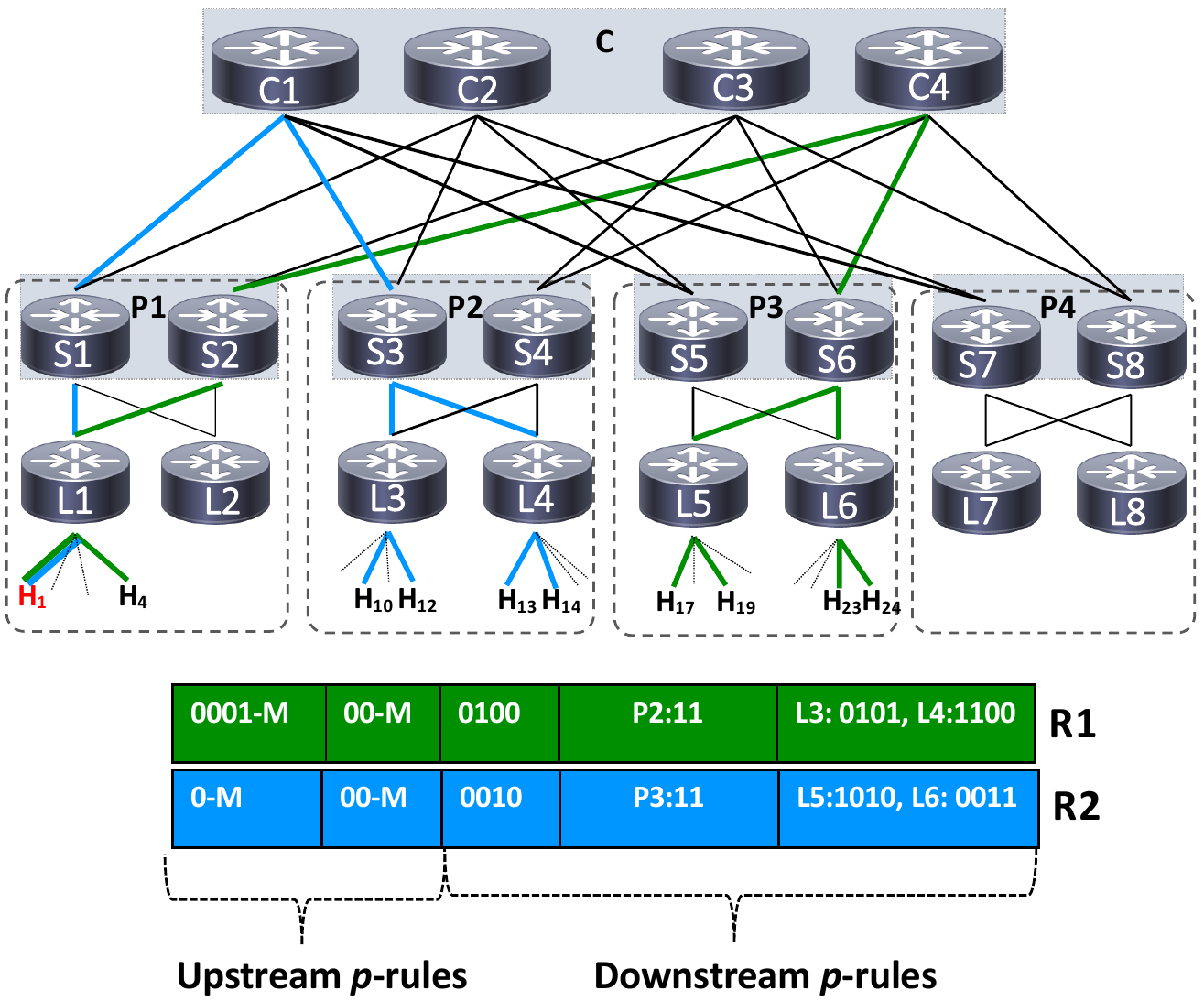}
   \caption{Locality aware clustering }
   \label{fig:fig2a}
\end{subfigure}%
\begin{subfigure}[b]{0.55\textwidth}
   \centering
   \hspace*{-1cm}
   \includegraphics[height=7cm, width=8cm]{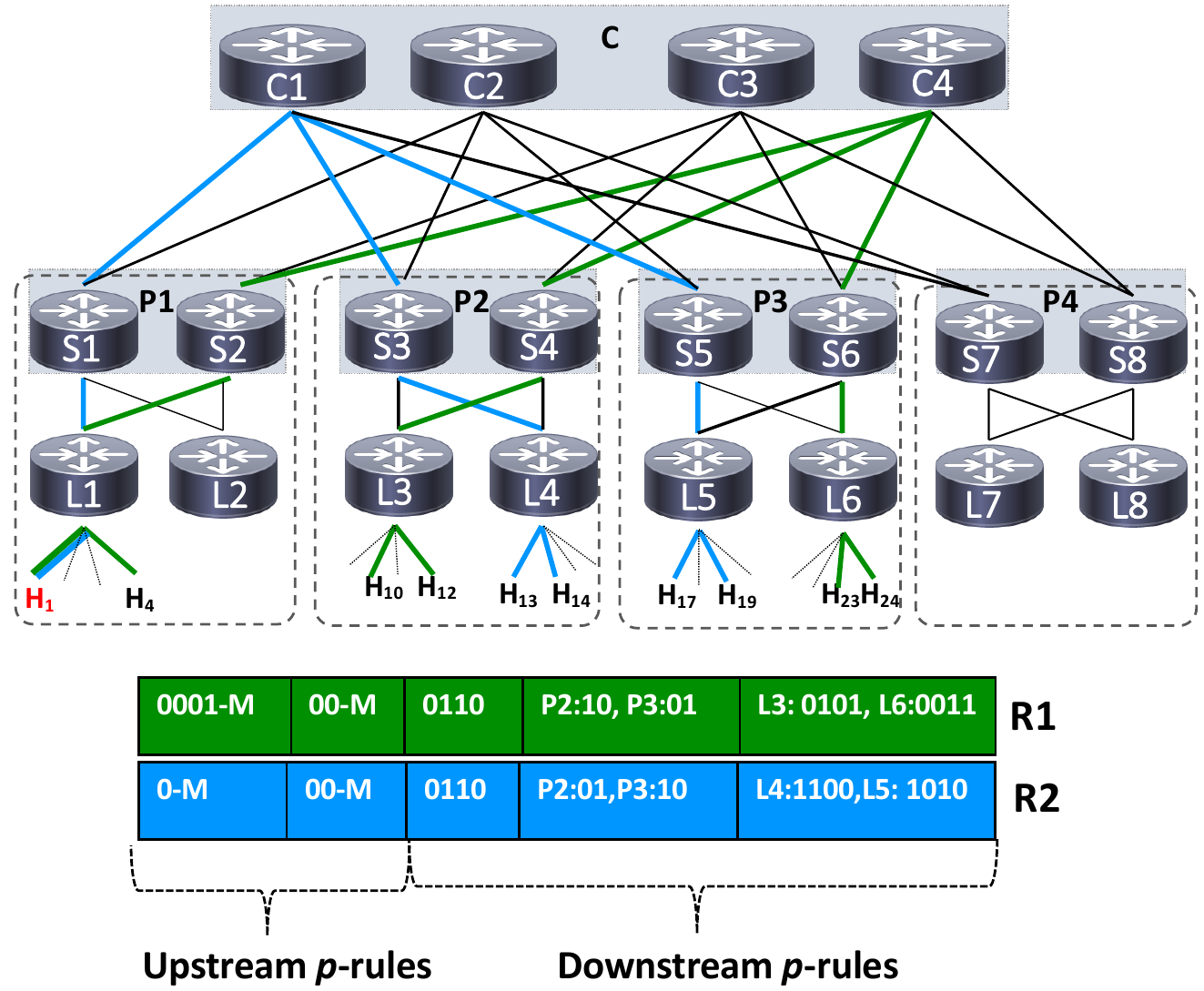}
   \caption{Locality oblivious clustering}
   \label{fig:fig2b}
\end{subfigure}%
   \caption{Clustering choice example of multicast tree on a three-tier Clos topology with four pods. In this topology, there are 4 hosts under each leaf switch (Top of Rack). $H_1$ is the source of multicast groups, and $H_4$, $H_{10}$, $H_{12}$, $H_{13}$, $H_{14}$, $H_{17}$,$H_{19}$,$H_{23}$, and $H_{24}$ are the receivers of multicas group. }
   \label{fig:fig2}
\end{figure*}

\subsection{\proposed}\label{subsec:proposed}
\proposed~aims to reduce the control message traffic by reducing the number of extra transmissions that Elmo incurs in the downstream paths, as well as the size of the multicast packet header. As illustrated in the motivating example given in the previous section, \proposed~achieves this goal by clustering the set of group members into $k$ clusters. This is done for each multicast group independently.
Before presenting the clustering approach of \proposed, we introduce the following notations/parameters of the studied three-tier DC: throughout, let us denote the number of pods by $n$, the number of ports per-leaf switch by $l$, the number of leaf switches per pod by $m$.
Note that although in traditional fat-tree DC, $m=n/2$ and $l=n/2$, for the sake of keeping our technique applicable to any tree-based DC topologies, we use the general parameter notation.
%
%
Also, let $L_{g,i}^j$ be the $l$-bit binary vector, corresponding to the $j$th leaf switch belonging to the $i$th pod, where $1\leq i \leq n$ and $1\leq j \leq m$, with each bit corresponding to one port of the leaf switch and taking $1$ when the port is serving a member of the multicast group $g$ and $0$ otherwise. For each multicast group $g$ and each pod $i$, let $L_{g,i}$ be the concatenation of the $m$ $l$-bit vectors of the $m$ leaf switches belonging to pod $i$.
That is, $L_{g,i}= L_{g,i}^1 || L_{g,i}^2 || ... || L_{g,i}^m$; here, $L_{g,i}$ is a binary vector of size $l\times m$.


%

Back to \proposed's clustering method, we begin by mentioning that in \proposed, we choose to cluster group members based on the pods as opposed to the leaf switches. That is, for each multicast group $g$, \proposed~clusters the set of $n$ vectors, $L_{g,i}$ with $1\leq i \leq n$, as opposed to the set $n\times m$ of vectors, $L_{g,i}^j$ with $1\leq i \leq n$ and $1\leq j \leq m$. This choice is supported shortly via an example.
\proposed~uses K-Means clustering algorithm with the Hamming distance as the distance metric, where the Hamming distance between two binary vectors is simply the number of bit positions in which they differ.
For each multicast group $g$, K-Means algorithm takes as an input the set of $n$ vectors, $L_{g,i}$ with $1\leq i \leq n$, and the number of clusters, $k$, and outputs $k$ clusters, with each cluster specifying a subset of the pods that need to belong to the same cluster.
Once clustering is done, the $p$-rules of each cluster are created by the hypervisor, which makes one copy of the multicast packet (data + header/$p$-rules) for each cluster.
For example, in Fig.~\ref{fig:fig1b}, when the hypervisor of host $H_1$ receives the multicast packet, it creates another copy of this packet, and adds the $R1$ rules to the first packet and the $R2$ rules to the second packet.

As with the case of K-Means clustering in general, the decision on the number of clusters, $k$, is a design choice, as the algorithm takes it as an input. The observation we made is that the larger the $k$ is, the lesser the number of extra transmissions in the downstream path and the lesser the header size overhead, but also the greater the number of extra transmissions in the upstream links (from leaf switches to core switches). However, we also observe that the overall (including both upstream and downstream paths) traffic overhead reduction improves with the number of clusters, $k$. More on this is provided in the evaluation section.

Now the reason for why \proposed~adopts clustering based on pods and not on leaf switches is as follows: if we cluster the downstream pods based on the $p$-rules for the downstream leaf switches regardless of which pod they belong to, extra packets transmissions will occur at the core and spine switches in the downstream path. For example, in Fig.~\ref{fig:fig2b}, when clustering is based on leaf switches only and when using the Hamming distance similarity, $L_4$ and $L_5$ will be clustered in the same cluster (e.g. $R2$), and $L_3$ and $L_6$ will be clustered in the other/second cluster (e.g. $R1$). In this case, because $L3$ and $L4$ are in the same pod (pod 2) but they are in different clusters, the packet will be sent twice at both core and spine downstream layers. The same thing happens with $L_5$ and $L_6$. To avoid this, \proposed~adopts a clustering choice that is locality aware of leaf switches (see Fig.~\ref{fig:fig2a}).

\subsection{Key Features of \proposed}
\label{subsec:features}
Compared to Elmo, \proposed~reduces multicast traffic substantially, and does so by:

\subsubsection{Reducing Packet Header Size}
In multi-rooted Clos topologies, unlike traffic load in upstream paths which are equally distributed, downstream paths are much heavier and are always the main bottleneck of the network. This is because,  in these types of topology, the upstream routing is fully adaptive, while the downstream routing is deterministic. Moreover, the multicast workload may make this worse because multicast packets are replicated at the downstream paths in order to reach each group member. In Elmo, by adding the $p$-rules to the packet, a data packet may have as many as 325 bytes of forwarding rules per each packet. In~\proposed, the average header size for the downstream packet is inversely proportional to the number of clusters $k$, i.e., $\frac{1}{k}$, of that of Elmo's packet, as explained in the previous subsection.

\subsubsection{Reducing Number of Extra Transmissions}
\proposed~first clusters the multicast group members into $k$ clusters, and then sends one copy (with same payload but different rules/header) for each cluster in the downstream, thereby reduceing the number of extra transmissions substantially in the downstream path.



%
%
%
\section{\sc {Performance Evaluation}} \label{sec:Evaluation}
Using simulations, in this section, we evaluate and compare the performance of \proposed~to that of Elmo in terms of their ability to reduce multicast control traffic. Mimicking the experiment setup of Elmo~\cite{Elmo}, we simulate a 3-tiered DC topology built with $48$-port switches, all of which connecting 27,648 servers, while considering different multicast group sizes. Group members for each simulated multicast group are distributed (uniformly) randomly across the servers. Let $l=48$ denote the number of ports in each of the leaf switches.

\subsection{Extra Packet Transmission Overhead} \label{subsec:ET}
We focus on downstream leaf switches here, since they are mostly the ones that cause extra packet transmissions. In this evaluation, we impose only one $p$-rule per packet for all the downstream leaf switches.
For Elmo, this one rule, denoted by $M$, is constructed as a bitwise-OR of $l$-bit vectors of all leaf switches that happen to be hosting at least one group member.
That is, $M = OR \{L^j_{g,i}\}_{1\leq i \leq n, 1\leq j \leq m}$.
%
For \proposed, one rule $M^u$ is to be constructed for each cluster $u$, $1\leq u \leq k$, also by bitwise-ORing all $l$-bit vectors of all leaf switches that happen to be hosting at least one group member and whose pod happens to belong to cluster $u$. 

The number of extra packet transmissions $ET$ incurred by Elmo~and \proposed~can be calculated as the sum of the Hamming distances/XOR between the rule and each of the $l$-bit vector of each leaf switches participating in the multicast tree/group. 
That is, for multicast group $g$,
$$ET_g^{Elmo} = \sum_{i \in S_g} XOR (L_i,M)$$
where $S_g$ is the set of all leaf switches hosting at least one member of mulitcast group $g$, and $L_i$ is the $l$-bit binary vector of leaf switch $i$. Similarly, 
$$ET_g^{\proposed} = \sum_{u=1}^{k}\sum_{i \in S^u_g} XOR (L_i,M^u)$$
where $S^u_g$ is the set of all leaf switches whose pods belong to cluster $u$ and that are hosting at least one member of mulitcast group $g$, and $k$ is the number of clusters per multicast group.

\begin{figure}
\centering
     \includegraphics[width=\linewidth, height = 5cm]{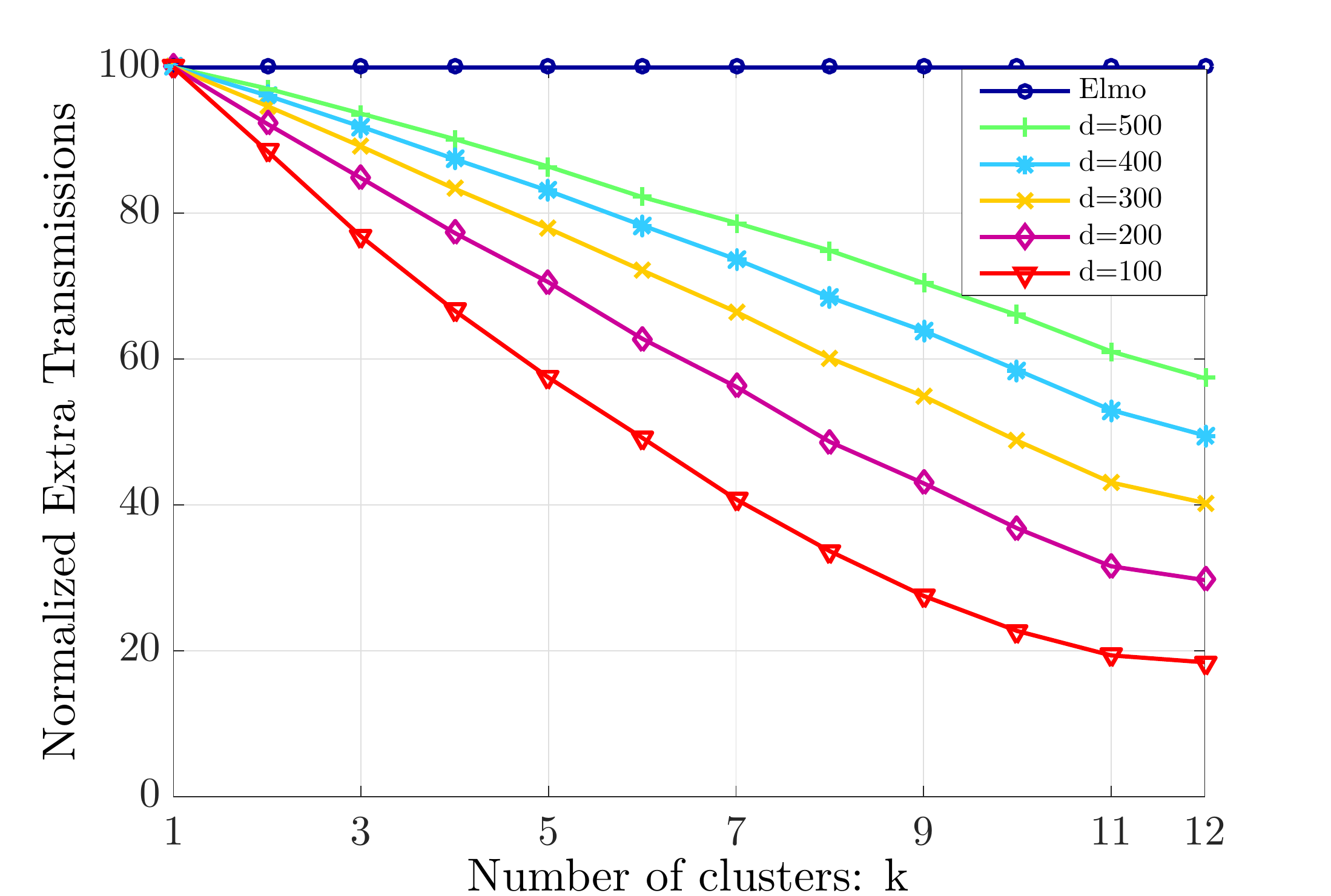}
      \caption{Number of extra transmissions caused when combining $p$-rules}
      \label{fig:3}
   \end{figure}

\begin{figure}
\centering
     \includegraphics[width=\linewidth, height = 5cm]{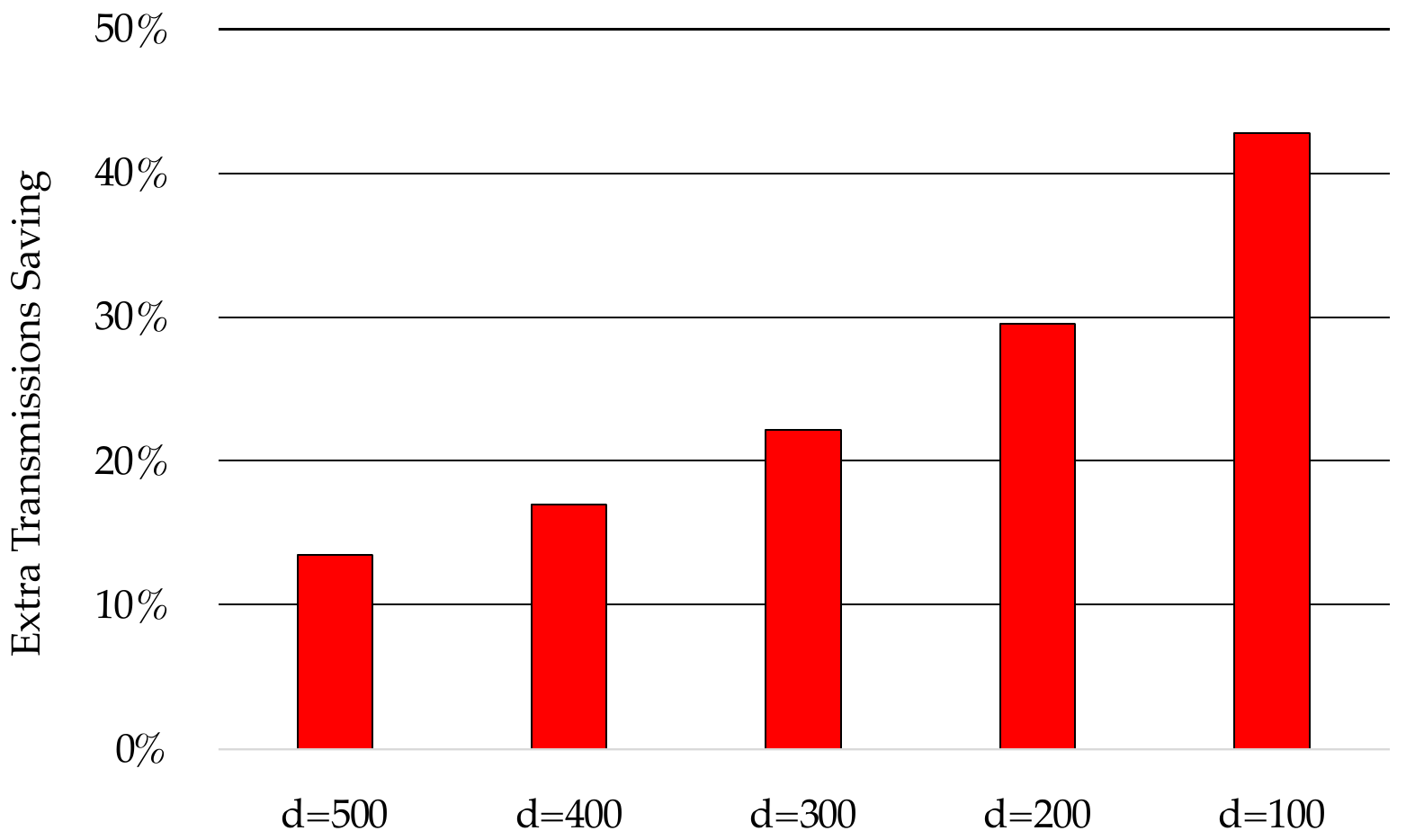}
      \caption{Extra transmissions saving for different size of a multicast group, when k=5 }
      \label{fig:4}
   \end{figure}

We vary the size of the multicast group from $d=100$ to $d=500$ members. Fig.~\ref{fig:3} shows the total number of extra packet transmissions for all the downstream leaf in the multicast tree caused by combining their $p$-rules. From Fig.~\ref{fig:3}, we observe that in \proposed, the number of extra transmissions depends on the size of the group as well as on the number of clusters. First, observe that \proposed~reduces the number of extra packet transmissions when compared to Elmo, especially when the number of cluster is increased.
For example, \proposed~reduces the number of extra transmissions from $10\%$ to $70\%$ for  group size of 200 members when the number of clusters is increased from $k=2$ to $k=12$.
The second observation is that the reduction of the number extra packet transmissions in \proposed~increases when the multicast group size is decreased. For example, Fig.~\ref{fig:4} shows that for $k=5$, extra packet transmissions decreases from about $43\%$  to about $14\%$ when the group size increases from 100 to 500 members. Here all numbers are normalized with respect to the total number of extra transmissions incurred by Elmo.


   \begin{figure}
\centering
     \includegraphics[width=\linewidth, height = 5cm]{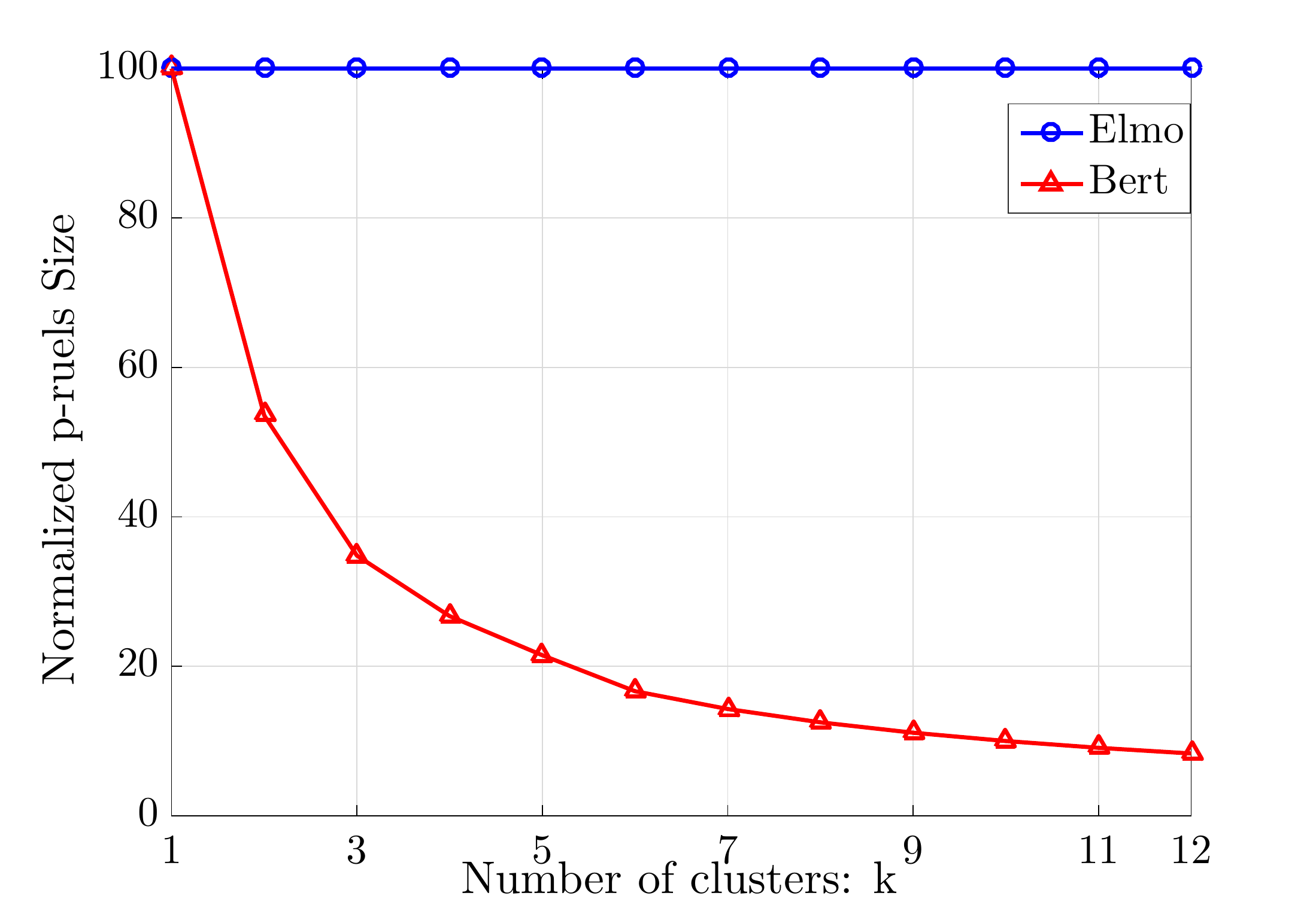}
      \caption{Header size of $p$-rules that sent to all downstream switches of a multicast group.}
      \label{fig:5}
   \end{figure}

\subsection{Header Size Overhead}  \label{subsec:Traffic}
Again, we focus on the downstream leaf switches because they use up most of the forwarding header capacity. In this experiment, we focused on multicast groups with large sizes; e.g. 2000 members. 
Fig.~\ref{fig:5} shows that the header size is dramatically decreased in \proposed, especially when the number of cluster is small. For example, when $k=2$, size of the header is reduced by $54\%$. Moreover, it gently keeps decreasing when the number of cluster is increased.

\begin{figure}
\begin{center}
\begin{minipage}[b]{1.0\linewidth}
\includegraphics[width=\textwidth, height = 4cm]{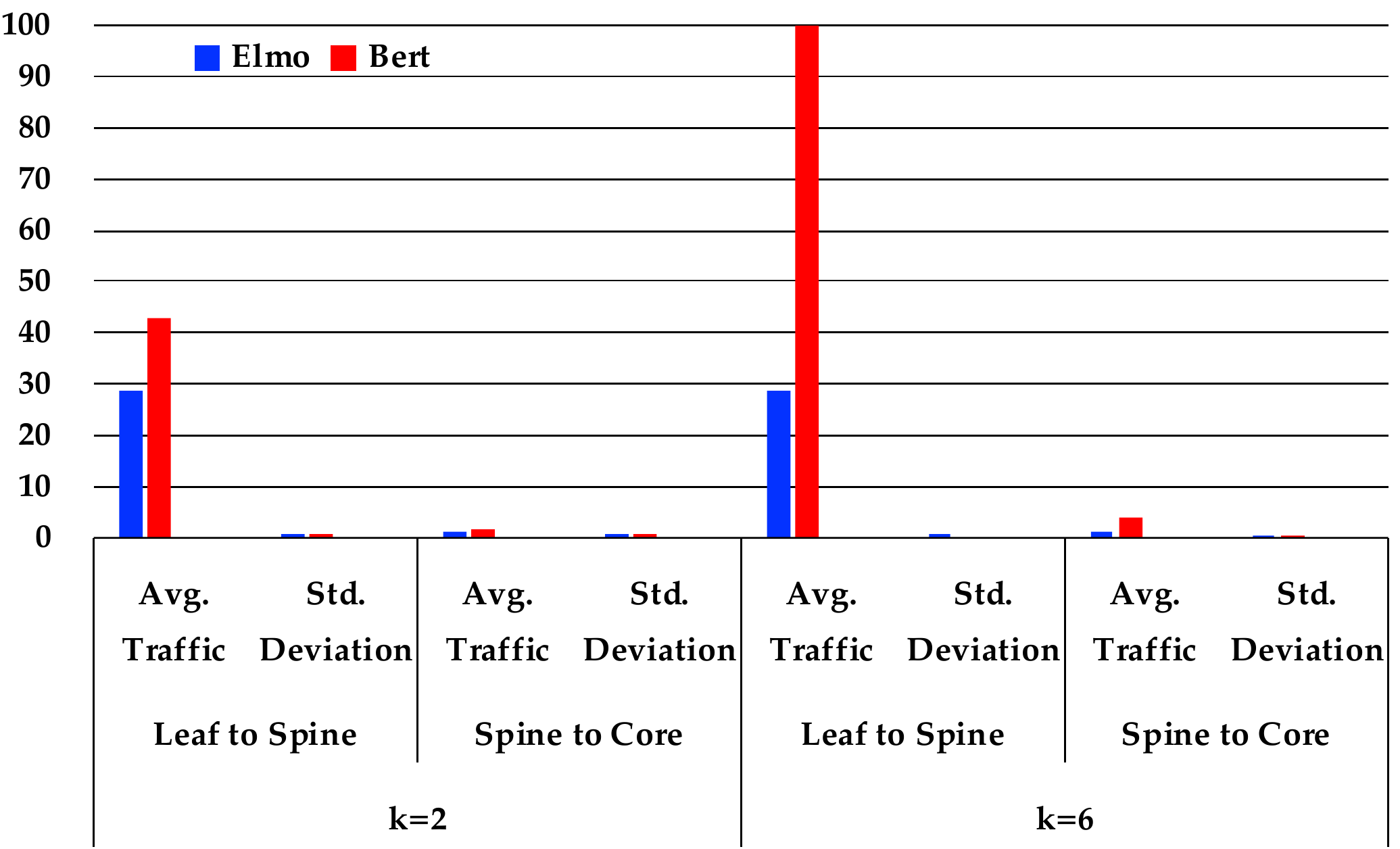}
\caption{Link load traffic in the upstream paths caused by a multicast flow of size 1000 pkt and d=2000.}
\label{fig:6}
\end{minipage}
\end{center}
\vspace{0.3cm}
\begin{center}
\begin{minipage}[b]{1.0\linewidth}
\includegraphics[width=\textwidth, height = 4cm]{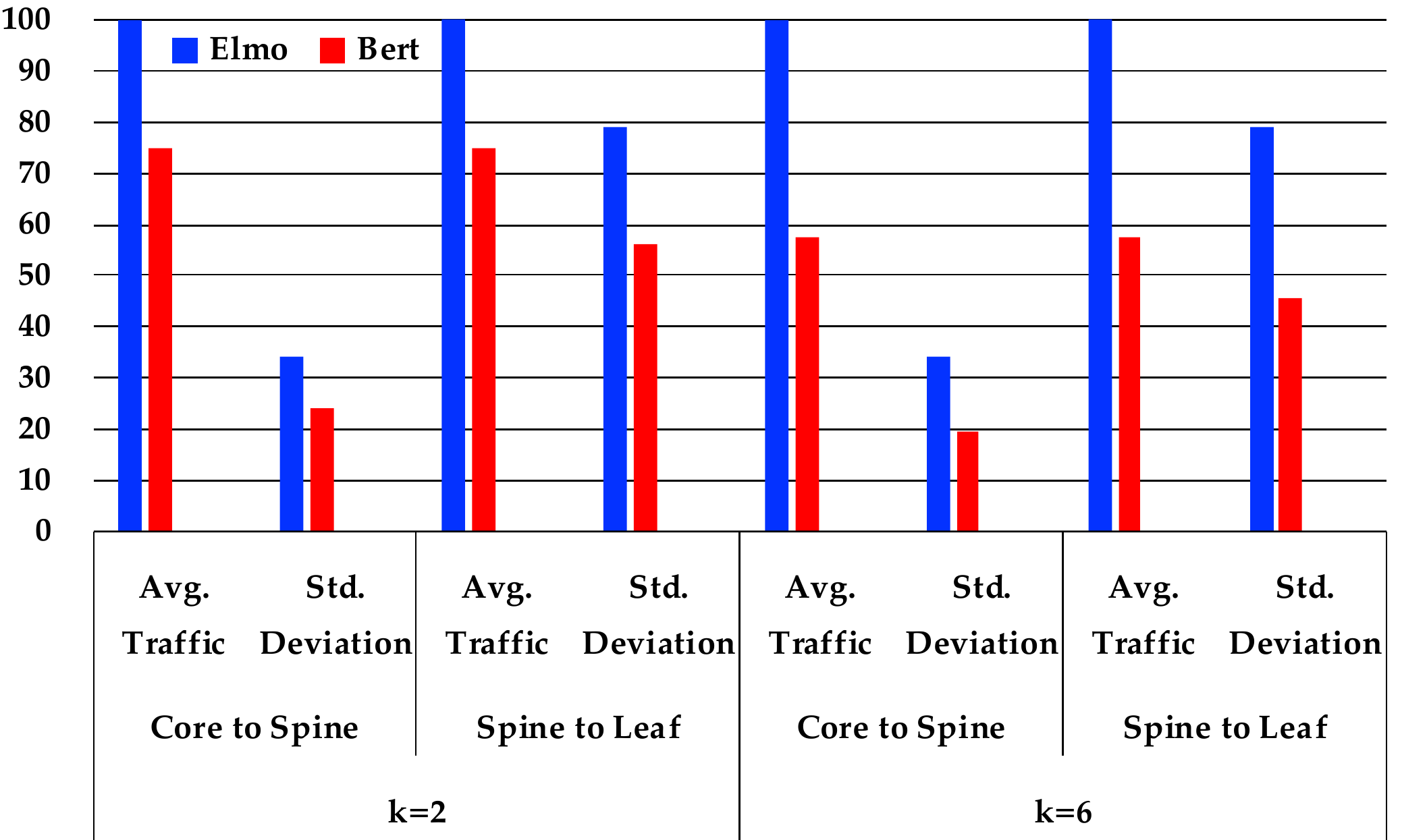}
\caption{Link load traffic in the downstream paths caused by a multicast flow of size $1000$ packets. Multicast group size is $d=2000$}
\label{fig:7}
\end{minipage}
\end{center}
\end{figure}

Now, in Figs.~\ref{fig:6} and~\ref{fig:7}, we show the impact of the header size reduction as well as the packet duplication caused by \proposed's proposed clustering. We calculate the average traffic traversing each link on upstream and downstream paths of each layer. Without loss of generality, we assume that the size of both the forwarding header and payload of the packet is one unit traffic each, and consider the multicast flow size for this group to be 1000 packets. We also assume that Equal-Cost Multipath protocol (ECMP)~\cite{ECMP} is used for load balancing traffic, and use the standard deviation across all links' traffic loads to show the evenness of load distribution across the links in each layer. In this experiment, we show the tradeoffs between large and small values of $k$ discussed in Sec.~\ref{subsec:proposed}.

%

Fig.~\ref{fig:6} shows that the average link traffic load in the upstream path achieved under \proposed~is higher than that obtained under Elmo. This is expected because \proposed~creates and sends multiple packets one for each cluster. For example, when $k=2$,  the upstream traffic load under \proposed~is $34\%$ higher than than under Elmo. Moreover, this traffic load increases when the number of clusters (e.g. when $k=6$). However, these links are evenly utilized as shown via standard deviation values.

On the other hand, in the downstream path, shown in Fig.~\ref{fig:7}, \proposed~achieves lower link traffic loads compared to Elmo, and this is true regardless of the number of clusters, though the more cluster \proposed~has, the lower the load. For example when $k=6$, the average link traffic observed under \proposed~is about $45\%$ lower than that observed under Elmo.

To sum up, when accounting for both the upstream and downstream paths, \proposed~outperforms Elmo also in achieving lower traffic loads across the links, leading to lesser network congestion.


\section{\sc {Conclusion}} \label{sec:Conclusion}
We proposed \proposed, a salable, source routed multicast scheme for cloud data centers. \proposed~builds on existing approaches to better suit nowadays cloud data center networks. \proposed~alleviates traffic congestion at downstream paths (usually highly congested links) by reducing both the packet header sizes and the number of extra packet transmissions. 

\bibliographystyle{IEEEtran}
\bibliography{References}
\end{document}